\documentstyle[pre,aps,tighten,multicol,epsfig]{revtex}

\begin{document}
\title{Thermal Treatment of the Minority Game}
\author{E. Burgos$^1$, Horacio Ceva$^1$\thanks{
e-mail ceva@cnea.gov.ar}, R.P.J. Perazzo${^{2,3}}$}
\address{$^1$Departamento de F{\'{\i}}sica, Comisi{\'o}n Nacional de\\
Energ{\'\i }a At{\'o}mica,\\
Avda. del Libertador 8250,1429 Buenos Aires, Argentina \\
$^2$Departamento de F{\'{\i}}sica FCEN,Universidad de Buenos Aires,\\
Ciudad Universitaria - Pabell{\'o}n 1, 1428 Buenos Aires, Argentina\\
$^3$Centro de Estudios Avanzados, Universidad de Buenos Aires,\\
Uriburu 950, 1114 Buenos Aires, Argentina}
\maketitle

\begin{abstract} We study a cost function for the aggregate behavior of
all the agents involved in the Minority Game (MG) or the Bar Attendance
Model (BAM).  The cost function allows to define a deterministic,
synchronous dynamics that yields results that have the main relevant
features than those of the probabilistic, sequential dynamics used for the
MG or the BAM.  We define a temperature through a Langevin approach in
terms of the fluctuations of the average attendance.  We prove that the
cost function is an extensive quantity that can play the role of an
internal energy of the many agent system while the temperature so defined
is an intensive parameter.  We compare the results of the thermal
perturbation to the deterministic dynamics and prove that they agree with
those obtained with the MG or BAM in the limit of very low temperature.

\end{abstract}

\begin{multicols}{2}

\narrowtext

\tightenlines

\section{Introduction}

The Bar Attendance Model (BAM) \cite{Arthur2} and the Minority Game (MG)
(see Refs.\cite{Challet} - \cite{Johnson}) have recently became regular
testing grounds to investigate how the individual actions of a system of
independent agents give rise to some kind of macroscopic ordering.  In the
MG, the agents have to make a binary decision which for the sake of
concreteness, it is usually taken to be associated to going or not going
to a bar.  The winning option is that of the minority.  The MG is a
particular case of the BAM which has in turn been introduced to show how
an ensemble of agents that perform inductive reasoning can self organize
to match some condition that is generally accepted to be the most
adequate.  In the case of the BAM this corresponds to the largest
acceptable attendance without incurring in some discomfort.

Both models have been compared with each other in Refs.\cite{bcp} and
\cite{bcp-II} working out a generalized version of the MG (the GMG)in
order to consider situations in which the minority is replaced by {\em an
arbitrary fraction} $\mu$ of the ensemble of players.  This is fixed
externally as a control parameter.  In all these models the players update
their attendance probabilities with a random correction, depending upon
the past record of successes and failures.  Asymptotic stable
configurations are always reached.  These are, however, of quite different
nature depending upon the values of the control parameters, of the initial
conditions and on the updating rules involved in each model.

In the present work we are interested in the cases in which the asymptotic
stable distribution can be assimilated to a kind of thermodynamic
equilibrium.  In these situations the agents continue to update their
attendance probabilities but the corresponding probability density
distribution remains stationary.  The stochastic dynamics that has been
developed for the BAM in ref.\cite{bcp-II} always leads the system to
these type of configurations while in the cases studied for the GMG, when
$\mu $ is significantly larger (or smaller) than 1/2, the system gets
stuck in quenched configurations that strongly depend upon the initial
conditions.  Updating stops because agents have accumulated a great number
of successes.  However, these ``glassy" states can nevertheless be
``melted" into equilibrium if the memory of past successes is repeatedly
eliminated in an iterative process that can be assimilated to an annealing
procedure.

A remarkable result that has been obtained in all numerical
simulations is that the equilibrium configuration entails a
diversity in the individual actions. The population is
drastically partitioned into two subsets, one that always goes to
the bar and the other that never goes. It therefore seems that in
spite of the fact that the agents do not exchange information,
they manage to coordinate their actions to proceed in two
opposite ways. The number of agents in both subsets are in a ratio
that is equal to $\mu/(1-\mu)$. Such polarization is not an
intuitive result. A na{\"{\i}}ve guess is to assume that all
agents should choose the same probability of attendance and this
should be equal to $\mu$. However this turns out to be not a
stable distribution because parties that are larger or smaller
than the accepted crowding occur with a great chance.

The fact that all agents adjust their attendance probabilities in order to
minimize their failures (i.e. to go when the bar is crowded or not go when
the bar is empty) leads to an aggregate behavior that minimizes a global
cost associated to inadequate attendances.  We propose to express such
cost by the second moment of the attendance with respect to the acceptable
level $\mu$.

The purpose of the present paper is to investigate the effects of
introducing that cost function in the relaxation dynamics of the
system. We show that this is a Lyapunov function for the many
agent system, i.e. it is possible to derive a deterministic
dynamics as the descent along its gradient, that monotonically
reduces its value. This corresponds to a heavily coordinated,
synchronous evolution.

We prove that the cost function meets the requirements of an internal
energy of the many agent system.  We also introduce a temperature
parameter through a Langevin-like approach that can be defined in terms of
the fluctuations of the attendance strategies.  Except for finite size
effects this can be proven to be an intensive parameter.  We also
superimpose thermal fluctuations to the deterministic dynamics mentioned
above.  Depending upon the amplitude of these fluctuations, the
polarization is gradually smeared until a point in which completely
disappears.

The thermally modified, relaxation process that we define here is
completely different from those involved in the  GMG or BAM
approaches that involve the independent and uncoordinated actions
of all the agents. The latter involves a random updating of
individual attendance strategies governed by a (small)
uncertainty amplitude that is interpreted as the precision of
such updating. We prove that in the limit of low temperature, and
small uncertainty amplitude both dynamics lead to entirely
equivalent asymptotic equilibrium configurations. The thermal
interpretation of the uncertainty amplitude also allows to cast
the annealing process presented in Refs.\cite{bcp} and
\cite{bcp-II} into a thermal framework as the well known case of
simulated annealing \cite{gelatt}.

In section II we derive the cost function, and in section III we
investigate the dynamics that corresponds to the descent along its
gradient.  In section IV we present a Langevin approach to define the
temperature in terms of the fluctuations that are present in the
asymptotic equilibrium configuration.  In Sec. V we compare this with more
traditional approaches for the relaxation process.  In section VI we draw
the conclusions.

\section{The cost function}

Consider a set of $N$ agents that have a probability $p_i (i=1,2,\dots,
N)$ to go to the bar.  The distribution of the $p_i$'s is given by the
probability density function $P(p)$.  As we shall shortly explain the
$p_i$ are updated in time according to some dynamics and therefore the
function $P(p)$ also changes in time.

In the ordinary rules of the GMG when a player goes to the bar and finds
it is crowded or when she does not go and the bar is empty, loses a point.
If the opposite happens she gains a point.  The level of crowding is
specified by the value of the control parameter $\mu$.  When her account
of points falls below zero she updates her attendance probability choosing
at random a different value within the interval $(p_i-\delta
p/2,p_i+\delta p/2)$.  When equilibrium is reached, the resulting
distribution $P(p)$ concentrates the population in the immediate
neighborhood of $p\simeq 0$ and $p\simeq 1,$ plus an almost vanishing
contribution from intermediate values.  The ratio of the areas below these
two peaks is close to $\mu /(1-\mu )$.

The aggregate behavior is associated to the density distribution ${\cal P}
(A)$ that gives the probability of occurrence of a party of $A$ customers
attending the bar.  The function ${\cal P} (A)$ is of course completely
determined by $P(p)$.  In order to calculate it let us assume without loss
of generality that all the agents distribute themselves into $D+1$
different bins of $n_d (d=0,1,\dots, D)$ agents each, with strategies $p_d
= d/D$.  The density distribution $P(p)$ can then be written as:

\begin{equation}
P(p)=\sum_{d=0}^D \frac{n_d}{N}\delta(p-p_d).
\end{equation}

\noindent With this assumption, the distribution ${\cal P}(A)$ can be
written as:

\begin{eqnarray}
{\cal P}(A)&=& \sum_{\ell_0=0}^{n_0} \dots
\sum_{\ell_D=0}^{n_D}\prod_{d=0}^D \lbrack {n_d \atopwithdelims()
\ell_d}p_d^{\ell_d}(1-p_d)^{n_d-\ell_d} \rbrack \nonumber \\
& & \times \delta(A-\sum_{d=0}^D\ell_d).
\label{distrigrupo}
\end{eqnarray}

We define the cost function for the whole ensemble of agents as in
ref.\cite{bcp-II}, namely as the second moment \cite{definicion}
with respect to the tolerated crowding level $\mu$:

\begin{equation}
{\cal C} = \sum_{A=0}^N (A-N\mu)^2{\cal P}(A)
\label{costo}
\end{equation}

In order to calculate it, we introduce Eq. (\ref{distrigrupo}) into the
definition of Eq.(\ref{costo}) and perform first the summation over $A$
taking advantage of the $\delta(A-\sum_d^D\ell_d)$.  Once this is done,
one can perform the summations involved in each of the terms in which
$(N\mu-\sum\ell_d)^2$ splits down.  The summations over different $\ell$'s
decouple from each other and result either in a 1; or in $n_dp_d$; in
$n_d^2p_d^2+n_dp_d(1-p_d)$ or in $(n_dp_d)(n_{d'}p_{d'})$.  These terms
can be gathered again to yield:

\begin{eqnarray}
{\cal C} &=& (N\mu-\sum_{d=0}^D n_dp_d)^2+\sum_{d=0}^D n_dp_d(1-p_d)
\nonumber \noindent \\ &=& N^2(\mu-<p>)^2+N(<p>-<p^2>)
\label{costo2}
\end{eqnarray}

\noindent where $<p^m>$ stands for $\sum_p p^m P(p) = \sum_d p_d^m n_d/N$
for $m=1,2$.  The expression of ${\cal C}$ given in Eq. (\ref{costo2})
contains no assumption about the system being in equilibrium.  This is the
reason why ${\cal C}$ is proportional to $N^2$ instead of being
proportional to the size $N$ of the system, as befits to an extensive
magnitude.  The numerical simulations however indicate that in equilibrium
$<p> =\mu$ and therefore this term cancels except for possible
fluctuations.  Actually the $O(N^2)$ term is eliminated by any
distribution $P(p)$ whose mean has the required value $\mu$.  For an
initial condition with uniformly distributed $p_i$'s and $P_o(p)=1/N$, as
it is used for most simulations, the cost is ${\cal
C}=N^2(\mu-1/2)^2+N/6$.  Such initial condition is a good guess for the
final distribution when $\mu \simeq 1/2$ (as for the most traditional
settings of the MG), but it is indeed very poor for the GMG when $\mu
\not= 1/2$.  In the next sections we discuss in greater detail the value
of ${\cal C}$ in equilibrium.

The na{\"{\i}}ve guess $P(p)=\delta(p-\mu)$ is also seen to cancel the
$O(N^2)$ terms in ${\cal C}$.  However such distribution causes that
parties with $A$ close to, but different from $N\mu$ occur with a sizable
probability.  The $O(N)$ in ${\cal C}$ are minimized precisely when the
probability of occurrence of such parties tends to zero by polarizing the
population into two subsets with opposite attendance strategies.  To see
this we approximate the two peaked equilibrium distribution that is
usually obtained in numerical simulations by

\begin{equation}
P(p)=\frac{n_1}{N}\delta(p-p_1)+\frac{n_2}{N}\delta(p-p_2)
\end{equation}

\noindent One readily sees that the $O(N^2)$ terms are eliminated when
$n_1p_1+n_2p_2=\mu N$ and the $O(N)$ terms are also eliminated if the two
peaks are $p_1=0;n_1=N(1-\mu)$ and $p_2=1; n_2=\mu N$.  The relaxation
dynamics that tends to minimize individual losses is therefore seen to
also optimize the global cost function defined in Eqs.  (\ref{costo}) and
(\ref{costo2})

\section{A deterministic dynamics for the GMG}

All the agents of the system, through uncoordinated actions minimize the
total cost ${\cal C}$ that is an aggregate function defined for the whole
system.  This fact suggests an alternative representation of the actions
of the agents as a synchronous, deterministic dynamics associated to the
descent along the gradient of ${\cal C}$.  This is described by the
following set of coupled differential equations for the $p_i$'s:

\begin{equation}
\frac{dp_i}{dt}=-\eta \frac{\partial{\cal C}}{\partial p_i}
         =\eta [2 N(\mu-<p>)-(1-2p_i)]
\label{motion}
\end{equation}

In Eq. (\ref{motion}) $\eta$ stands for a positive free parameter that -
as we shall shortly see - provides the scale for the time evolution of the
system.  The $O(N^2)$ and $O(N)$ terms in Eq.  (\ref{costo2}) are
translated into a fast and a slow dynamics that involve corrections of the
$p_i$ that are respectively $O(N)$ and $O(1)$.  To see this we first
derive the dynamics followed by $<p>$ by calculating the average over $i$
in both sides of Eq \ref{motion}.  We thus obtain:

\begin{equation}
\frac{d W(t)}{dt}=-2\eta (N-1)W(t)-2\eta(\frac{1}{2}-\mu)
\label{evolmedio}
\end{equation}

\noindent where we have set $W(t)=(<p>-\mu)$. This can explicitly
be integrated. The solution is:

\begin{equation}
W(t) = \frac{\mu-1/2}{N-1}+ W_o\mbox{e}^{-2\eta (N-1)t}
\label{valormedio}
\end{equation}

\noindent with $W_o$ standing for the initial value of $W(t)$.  This
expression allows in turn to find an approximate solution of the equations
of motion for the individual $p_i$'s.  To this end we write an asymptotic
approximation of Eq. (\ref{motion}) in which we assume that a long enough
time has elapsed so that $<p>-\mu$ can be approximated by the constant
term of $O(1/N)$ in Eq. (\ref{valormedio}).  By keeping only the leading
order in $N$ we obtain:

\begin{equation}
\frac{dp_i}{dt}= 2\eta(p_i-\mu).
\label{pindep}
\end{equation}

\noindent Note that dependence of $p_i(t)$ involves a {\em positive}
exponential.  However, this equation is not valid for $t\rightarrow
\infty$ because the fact that the $p_i$'s are probabilities, and are
therefore bounded between 0 and 1, it is not included in the equations but
rather in the boundary conditions of Eqs.  (\ref{motion}).

Eqs.(\ref{evolmedio}) and (\ref{pindep}) correspond respectively to the
fast and slow dynamics that have been mentioned above.  In the first place
we see that except for terms that are $O(1/N)$, $<p>$ approaches $\mu$
exponentially with the very short time constant $\lambda = 1/(2\eta N)$
that tends to zero as the system involves a larger number of individuals.
On the other hand, the differences $p_i(t)-\mu$ instead {\em grow}
exponentially for all $i$ indicating that the $p_i$'s {\em depart}
exponentially from the average $\mu$ and eventually saturate at its
largest or smallest possible values:  1 or 0, thus polarizing the
population of agents.  This process however takes place with a time
constant $1/(2\eta)$, that is $O(N)$ longer than the one involved in the
evolution of $<p>$ and is independent of the size of the system.  While
the average $<p>$ approaches very fast to the value $\mu$, the individual
$p_i$'s {\em depart} slowly from the same value.

Eqs.(\ref{motion}) can be tested numerically by
approximating them by finite differences. The individual attendance
probabilities $p_i$ are thus taken to be updated as
$p_i (t+1)= p_i(t)+\Delta (p_i)$ where :

\begin{equation}
\Delta (p_i)= \eta [2 N(\mu-<p>)-(1-2p_i)]
\label{dinamica}
\end{equation}

\begin{figure}[tbp]
\epsfig{file=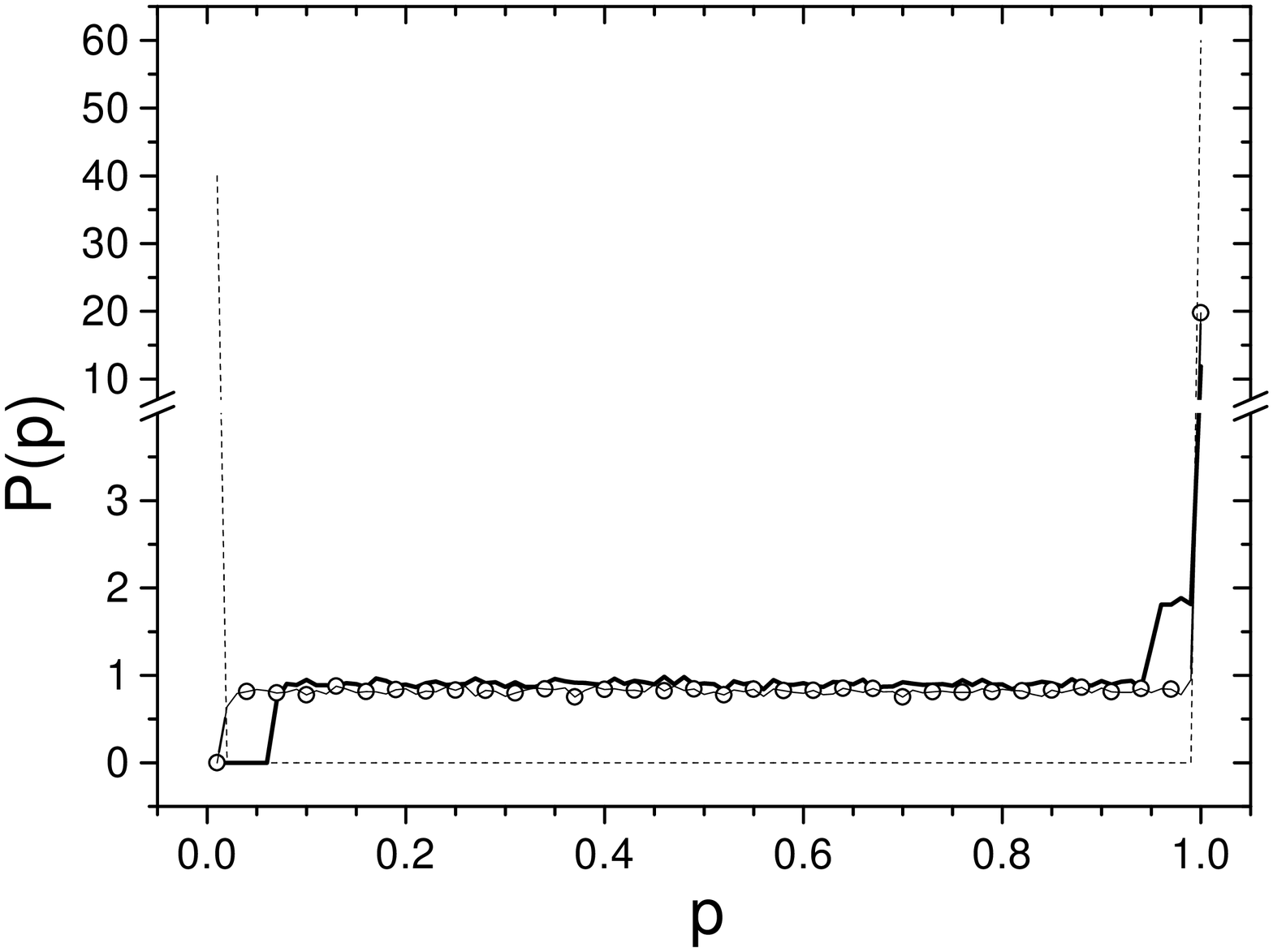,width=80mm}
\caption{Probability density distributions obtained after $10^4$
steps (solid line), $2\times 10^4$steps (open circles) and $2
\times 10^6$ steps (dash line),  using Eq.(\ref{dinamica}), $2\eta
N=1$, and $\mu = 0.6$. the first distribution shows a rigid
displacement to the right; the next ones show how the population
is progressively polarized.} \label{minimizacion}
\end{figure}

The resulting density distributions $P(p)$ that are obtained with this
dynamics are shown in Fig. \ref{minimizacion}.  The value of $\eta$ and
therefore that of the time constant $\lambda$ is in principle arbitrary.
However if $\lambda \gg 1$ the only effects that are noticeable are those
of the fast dynamics while if $\lambda \ll 1$ the descent towards the
minimum keeps bouncing at opposite sides of the quadratic well and never
reaches its bottom.  When $1/2 \ {\small \stackrel{<}{\sim}} \ \lambda \
{\small \stackrel{<}{\sim}} \ 2$ the descent is gradual enough so that the
interplay of both terms in $\Delta (p_i)$ leads the system to a minimum of
${\cal C}$.


The intermediate stages in the gradient descent are also shown in
Fig. \ref{minimizacion}.  In the first few steps the (fast) uniform
correction of $O(N)$ is seen to shift rigidly the initial distribution to
one side with the aim of adjusting the value of $<p>$ to that of $\mu$.
As a consequence, agents are piled up in one end while the other is
completely cleared.  Once the leading term in ${\cal C}$ is nearly
canceled, the slow dynamics gradually gathers agents at {\em both} ends of
the distribution producing minor fluctuations in the value of $<p>$.  The
density distribution $P(p)$ that is finally obtained is seen to correspond
to a strongly polarized population thus reproducing the main feature of
the equilibrium distributions obtained with the rules traditionally used
in the GMG or the BAM.

The present approach yields a density distribution that displays the same
polarization that is found in the GMG or in the BAM.  It is remarkable
that such a general qualitative agreement is found, although those
frameworks differ deeply from the deterministic formulation.  The
conceptual difference between the two approaches lies in the special role
played by the record of successes and failures that is kept in the BAM or
GMG and that is completely absent in the present treatment.  The usual
rules of the GMG can thus be considered to correspond to a dynamics
constrained by the (positive) balance of points that have been accumulated
in the past instances of the game.  There are other differences that
deserve further discussion.  These are related to the stochastic elements
of the dynamics used in that framework which are absent from the present
one.  Within this approach, these can be assimilated to the effects of a
finite temperature.  We turn to this point in the next section.

\section{Thermal fluctuations}

The usual rules of the BAM or the GMG involve a stochastic updating of the
attendance probabilities of each customer. When the account of points
of the $i-$th player falls below zero a new value of $p_i$ is chosen
{\em at random} from the interval $(p_i-\delta p/2,p_i+\delta p/2)$. This
can be interpreted as a kind of thermal fluctuation in which $\delta p$
can be related to the temperature.

A few qualitative features support this.  In equilibrium, the population
is drastically polarized into those that consistently go to the bar (and
therefore $p_i=1$) and those that do not go ($p_i=0$).  A small fraction
having $p_i$'s with intermediate values continuously migrate between both
extreme strategies.  This migration causes that the value of $<p>$
fluctuates around $\mu$.  These random values of $<p>$ have a distribution
that is sharply peaked at that value and has a width that is regulated by
$\delta p$.  In what regards the density distribution $P(p)$, a small
value of $\delta p$ produces sharp peaks at $p=0$ and $p=1$ and $P(p) \sim
0$ for intermediate values.  For larger values of $\delta p$ there is a
larger fraction of players that migrate between $p=0$ and $p=1$ thus
producing a rising in the ``bottom" of the distribution $P(p)$.

The above qualitative arguments provide hints to introduce thermal
fluctuations in the deterministic dynamics presented in the preceeding
section and also about their relationship with $\delta p$ for the case of
the GMG.  However a singular situation occurs for $\delta p \rightarrow 0$
that is associated to an infinitely long relaxation process or when
$\delta p > 1$ in which this parameter loses its physical meaning of a
being a probability.

Thermal-like fluctuations can formally be introduced following the same
steps as the Langevin approach to describe a Brownian particle.In the
present situation we start with the Eq.  (\ref{evolmedio}) for the motion
of the average value $<p>$, and we add a stochastic term $L(t)$ that
accounts for the random fluctuations

\begin{equation}
\frac{d W_s(t)}{dt}=-2\eta (N-1)W_s(t)-2\eta(\frac{1}{2}-\mu) + L(t)
\label{evolmedio2}
\end{equation}

\noindent We have added an index $s$ to $W(t)$ in Eq. (\ref{evolmedio})
to stress the fact that this is the value of $W(t)$ in the presence
of stochastic external fluctuations. The source of noise $L(t)$ can be
taken to be the average of $N$ uncorrelated sources of random fluctuations
affecting all the independent agents. One still has to specify a parameter
related to the statistical properties of the distribution of the stochastic
function $L(t)$. We will shortly prove that this is closely related to the
temperature. As usual we assume:

\begin{eqnarray}
\overline{L(t)} &=& 0  \label{media0} \\
\overline{L(t)L(t')} &=& \Gamma \delta(t-t') \label{gamma}
\end{eqnarray}

\noindent In Eqs. (\ref{media0}) and (\ref{gamma}) and in all
what follows $\overline{(\dots)}$ denotes an average over a
suitable ensemble of replicas of the $N-$ agent system. The
parameter $\Gamma$ is a constant that represents the mean square
amplitude of instantaneous, uncorrelated perturbations. The
stochastic differential equation (\ref{evolmedio2}) can
explicitly be integrated. The result is

\begin{equation}
W_s(t) = W(t)+
 \mbox{e}^{-2\eta (N-1)t}\int_0^t\mbox{e}^{2\eta (N-1)\omega}L(\omega)d\omega
\label{valormedio2}
\end{equation}

\noindent where $W(t)$ is the solution given in
Eq(\ref{valormedio}) in which no fluctuations are present. If an
average is made on both sides of Eq. (\ref{valormedio2}), over a
sub-ensemble of systems having the same initial conditions $W_o$
appearing in Eq. (\ref{valormedio}), one can immediately see that
Eq. (\ref{media0}) implies that $\overline{W_s(t)} = W(t)$ and
therefore the convergence of $<p>$ to $\mu$ (up to terms $O(1/N)$
is also insured within the stochastic dynamics. If the mean
square fluctuations of $W_s(t)$ are calculated with the aid of
Eq. (\ref{gamma}), we get:

\begin{equation}
\overline {W_s^2(t)}=W^2(t)+\frac{\Gamma}{4\eta
N}\quad\Bigl[1-\mbox{e}^{-4\eta N t}\Bigr]\quad
\label{fluctua}
\end{equation}

\noindent The effect of the stochastic term in $W_s(t)$ produces a non
vanishing value $\overline {W_s^2(\infty)}$.  In ordinary statistical
mechanics, the mean square fluctuations of the stationary solution of the
velocity of Brownian particles is directly related to its average kinetic
energy and can be set equal to $kT$.  By analogy we formally define a
temperature parameter $T$ that is independent from the size of the system,
as the mean square fluctuations of $<p>$ in an equilibrium configuration,
scaled by the number of agents of the system.  Neglecting terms $O(1/N^2)$
we obtain:

\begin{equation}
T \doteq N\overline {(<p>-\mu)^2}= \frac{\Gamma}{4\eta}
\label{temperatura}
\end{equation}

\noindent The parameter $\eta$ is a factor relating $T$ with the
amplitude of the random fluctuations and plays a similar role
than the Boltzmann constant.

Eq. (\ref{temperatura}) allows to write the ensemble average of the
cost $\overline{{\cal C}}$ for an equilibrium configuration
and for finite temperature. Up to the leading order in $N$ we obtain:

\begin{eqnarray}
\overline{{\cal C}}&=& N^2\overline{(\mu-<p>)^2}+N\overline{(<p>-<p^2>)}
\nonumber \\
                   &=& N[ T + \mu -\overline{<p^2>}]
\label{costotermico}
\end{eqnarray}

\noindent $\overline{{\cal C}}$ is a positive, extensive magnitude
which, in equilibrium, grows linearly with the size of the system and
can therefore be taken to play the role of an internal energy.

\begin{figure}[tbp]
\epsfig{file=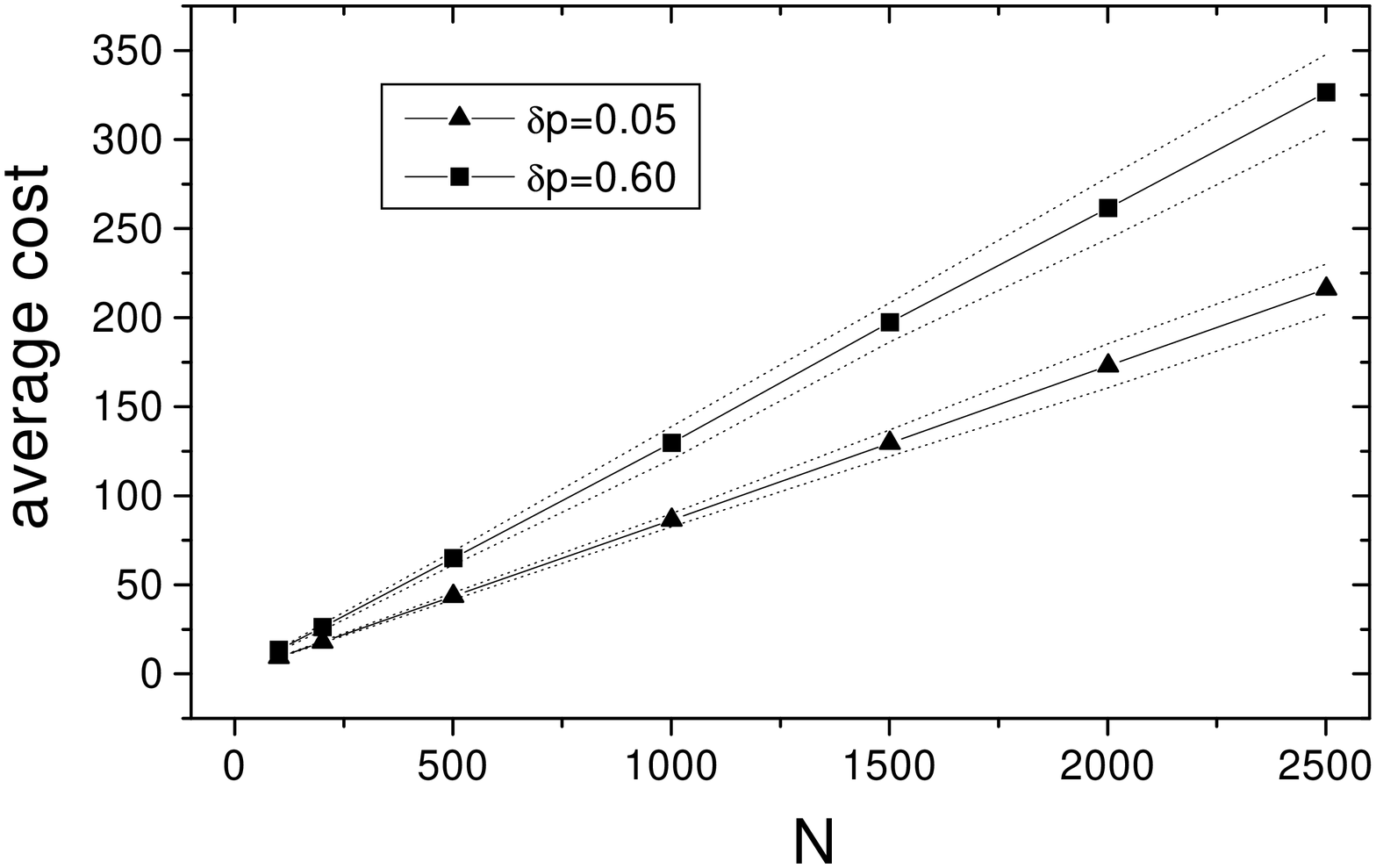,width=80mm}
\caption{Linear dependence of ${\overline{\cal C}}$ as a function of $N$
for the GMG, $\mu=0.6$ and different values of $\delta p$}
\label{costolineal}
\end{figure}

The linear dependence of  $\overline{{\cal C}}$ with the size of
the system can be checked for the GMG. To do so we have
calculated the cost using the definition of Eq.(\ref{costo}), with
different number of agents. We first allowed the system to relax
to the asymptotic equilibrium configuration and performed a
suitable ensemble average over several replicas of the system. The
linear dependence is shown in Fig. \ref{costolineal}. The last
iteration steps are used to estimate the dispersion of the
numerical result and is shown with a pair of dotted lines. The
slope of these lines change slightly with the parameter $\delta
p$ of the GMG. This is due the relation between $T$ and $\delta
p$ that we discuss later.


\section{Thermal relaxation}

To include thermal fluctuations into a numerical treatment of the
deterministic dynamics amounts only to introduce a random additive term in
Eq.(\ref{dinamica}), namely:

\begin{equation}
p_i(t+1)= p_i(t)+ \Delta (p_i)+L_{\tau}^{(i)}(t)
\label{termico}
\end{equation}

\noindent where $L_{\tau}=\tau(1/2-r)$ and $r$ is a random number
uniformly distributed in the interval $[0,1]$.  This function represents
the fluctuations produced on the $i$-th agent by a thermal bath.  The
temperature is defined by the second moment $\Gamma$ of the probability
density of the $L_{\tau}^{(i)}(t)$.

\begin{figure}[tbp]
\epsfig{file=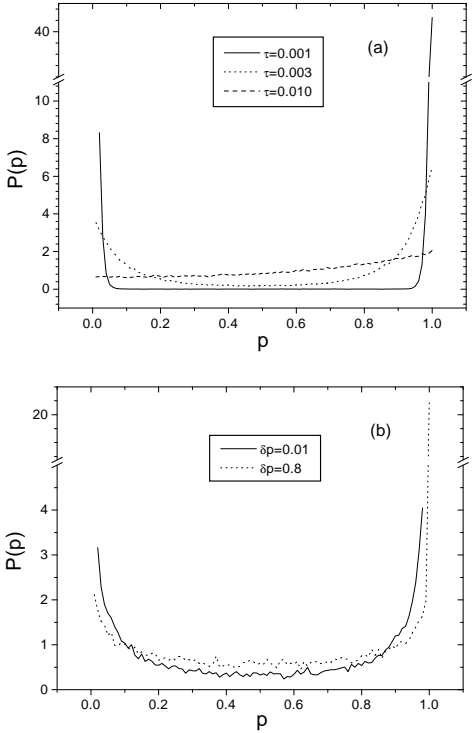,width=80mm}
\caption{{\em (a)} Probability density distribution obtained with the
thermal dynamic of Eq.(\ref{termico}) for the values of $\tau$ that are
shown in the inset. {\em (b)} Same distributions obtained with the
stochastic dynamics of the GMG, for the values of $\delta p$ shown in
the inset.}
\label{P(p)termica}
\end{figure}

The limit in which $L_{\tau}^{(i)}(t)$ has zero width (and therefore $\tau
=0$) corresponds to the deterministic dynamics discussed in Sec.  II.
Larger values of $\tau $ are associated to fluctuations that may
eventually override the updating amplitude $\Delta (p_i)$ and tend to
smear the distribution with two sharp $\delta$-functions, increasing the
fraction of the population that have strategies $p_i \not= 0$ or 1. (see
Fig. \ref{P(p)termica}$(a)$).  If $\tau$ is further increased the
polarization is progressively destroyed because the drift of the $p_i$'s
towards 0 or 1 has to equilibrate against random shocks that prevent them
to reach those limiting values.


\begin{figure}[tbp]
\epsfig{file=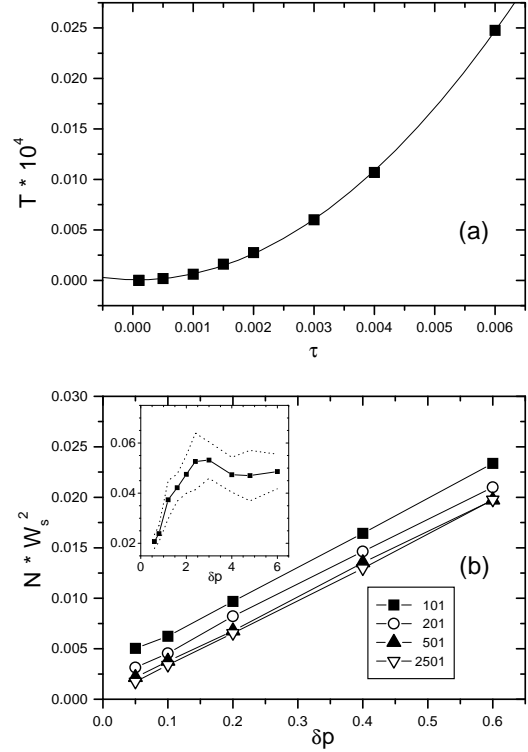,width=80mm}
\caption{{\em (a)} Relationship between $\tau$ and $T$ as defined
in Eq. (\ref{temperatura}). Solid squares correspond to the
numerical calculation, while the line is the quadratic regression
$10^4 T=5.43 10^{-5}-.114 \tau +703.9 \tau^2$, with $R^2=0.9999$.
{\em (b)} linear dependence of the fluctuations $N \overline
{W^2_s}$ with $\delta p < 1$ for the GMG, and several values of $N$
(indicated in the figure). The upper inset shows that
fluctuations saturate at a limiting value $ \sim .05 $ if the
plot is extended for $\delta p >1 $} \label{T_vs_tau}
\end{figure}

Given the stochastic dynamics of Eq.(\ref{termico}) together with the
definition in Eq.(\ref{temperatura}) it is possible to calculate the value
of $T$ in an equilibrium configuration, and relate $T$ with $\tau$.  The
parameter $\eta$ has to be chosen such that the relaxation of the
deterministic dynamics is guaranteed i.e. when the time constant $\lambda
= 1/(2\eta N)$ introduced in Sect. 3 is $\lambda \sim 1$.  In
Fig. \ref{T_vs_tau} we show that, as expected, $T\sim\tau^2$.

Eq.(\ref{temperatura}) allows also to calculate $T$ in any configuration
reached through the stochastic dynamics of the GMG or the BAM.  With this
we can check two important features.  The first is an estimation of the
finite size corrections in the definition of $T$ given in
Eq.(\ref{temperatura}), i.e. the regime in which $T$ is independent of the
size of the system.  The second outcome is to establish a quantitative
relationship between $T$ and $\delta p$ that goes into the relaxation
dynamics of the GMG.

We have calculated $W_s^2(t)$ for the GMG using several values of $\delta
p$ and $N$.  We have allowed $t$ to be large enough to reach equilibrium.
We have then performed an ensemble average over several replicas of the
system.  The last steps have been used to gauge the dispersion of the
numerical values.  The results are shown in Fig.\ref{T_vs_tau}$(b)$ where
we plot $N \overline{W_s^2(\infty)}$ as a function of $\delta p$.

All the above mentioned features can be extracted from Fig.
\ref{T_vs_tau}.  Firstly finite size effects are clearly seen to affect
only the smallest systems up to $N\sim 500$.  Second the independence of
$N \overline{W_s^2(\infty)}$ from the size of the system as assumed in the
definition of Eq.(\ref{temperatura}) follows from the fact that the curves
for $N \geq 500$ lump tightly together.  In the third place a linear
regression of all the curves establishes that $\delta p$ and $T$ have the
same physical interpretations, and within the interval considered are
nearly proportional to each other, namely $T =K \delta p$, with $K= (320
\pm 20) 10^{-4}$.


The fact that $T$ and $\delta p$ are conceptually equivalent leads to extend
the GMG simulations to higher values of $\delta p$. These values have seldom
been explored \cite{Ceva} in the literature because this parameter measures
the {\em minor} adjustments performed by the agents that try to find the
``best" attendance probability. Large values of $\delta p$ could for instance
correspond to irresolute or hesitating agents.

There are however important points that have to be considered.  In the
first place the value of $\delta p$ can not be taken arbitrarily large.
This is so because it measures the uncertainty of the value of a
probability.  Values of $\delta p \ {\small \stackrel{>}{\sim}} \ 1$ have
therefore little physical meaning.  In addition, if $\delta p$ is
nevertheless extended to values higher than 1 by any plausible analytical
extension (for instance using periodic or reflective boundary conditions),
the fluctuations $\overline {W_s^2}$ for $\delta p > 1$ are seen to
saturate at an approximately constant value  (see inset in Fig.
\ref{T_vs_tau}$(b)$).  These facts cause that the correspondence between
$\delta p $ and $T$ necessarily breaks down.

A comparison of the probability density distributions $P(p)$ obtained with
both approaches further supports this departure. In
Fig. \ref{P(p)termica}$(b)$ we show the equilibrium density distributions
that are obtained with the stochastic, asynchronous updating rules
of the GMG for two values of $\delta p$ (and $\mu=0.6$).
It is seen that these diverge from those of Fig.  \ref{P(p)termica}$(a)$
that are obtained with the dynamics given in Eq.(\ref{termico}). Note
however that there are noticeable ressemblances for small amplitude
fluctuations. See for instance the distributions plotted in full line in
Fig.\ref{P(p)termica}(b) and the one for $\tau=0.003$ in
Fig.\ref{P(p)termica}(a).

As mentioned before, the origin of the departure between both
dynamics can be found in the scoring of successes and failures
that is used in the GMG, that is absent in the present approach.
Some customers can be considered to be excluded from the updating
dynamics as a consequence of their great accumulation of points.
This, for instance, produces the large value of $P(p=1)$: many
players that have accumulated a large positive account attending
the bar do not change strategy. The scoring of each player works
as a kind of ``Maxwell Deamon" that classifies agents into
different groups, endowing each one with a different updating
rate.

The equilibrium configuration that is reached in the GMG therefore entails
a distribution of updating rates in which some players are essentially
frozen while others modify their attendance strategies frequently.
This situation is completely different to the one obtained with the
dynamics of Eq.(\ref{termico}) in which {\em all} agents undergo
stochastic perturbations in {\em every} time step.

\begin{figure}[tbp]
\epsfig{file=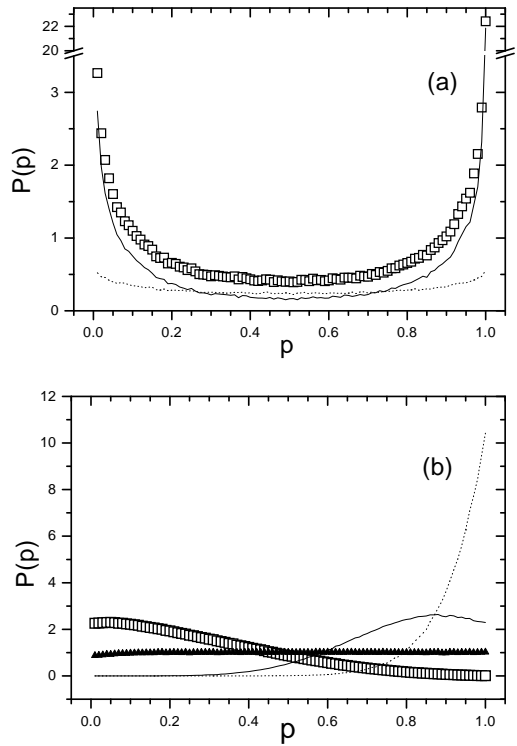,width=80mm}
\caption{Partial probability density distributions of individual
attendance strategies for the GMG for different subsets of
players obtained for 1001 players, crowding level of 600/1001,
and averages made over 2000 histories. {\em (a)} Asymptotic
distributions. Subset of players with more than 10 accumulated
points (full line) and with less than 11 points (dash line). The
total probability density distribution is shown with empty boxes.
{\em (b)} Density distributions at the end of the first 10 steps of
the simulation. Players with 0 points (open boxes) have the
greatest mobility, players with 5 and 10 points (full and dash
lines respectively) have lower mobility. The total density
distribution is shown in full triangles} \label{Maxwell}
\end{figure}

In order to show this we present in Fig. \ref{Maxwell} some
results of the GMG, in which we have used a large value of
$\delta p$ ($\delta p=0.8$) and we have arbitrarily partitioned
the ensemble of 1001 players into two sets. One of the sets
gathers all players having at most 10 points the other contains
all the rest. We have plotted their respective density
distributions $P(p)$. The agents having less that 11 points are
the ones that participate more strongly in the dynamics because
undergo more frequent updatings.


The above comparison indicates that the GMG and the thermal relaxation
dynamics of Eq.(\ref{termico}) strictly coincide only in the limit of $T
\rightarrow 0$.  However the strong qualitative resemblance of the results
for $\delta p \leq .6$ allows to interpret $\delta p $, with these
limitations, as equivalent to a thermal fluctuation.

\begin{figure}[tbp]
\epsfig{file=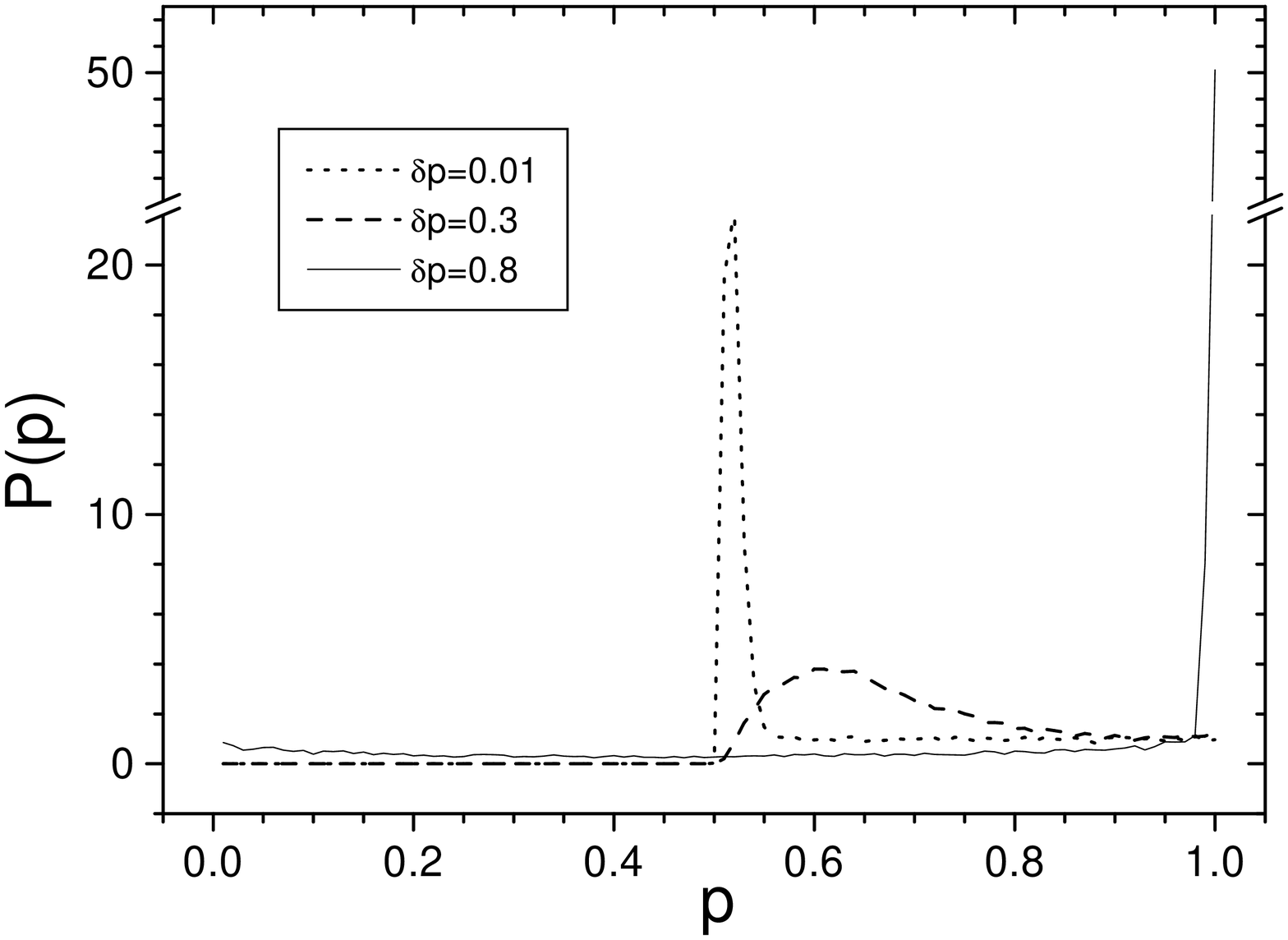,width=80mm}
\caption{Asymptotic probability density distributions of
individual attendance strategies for the GMG obtained with the
values of $\delta p$ that are shown in the inset. Notice that for
the highest value of $\delta p$ there is no quenching.}
\label{melting}
\end{figure}

The thermal interpretation of $\delta p$ has one interesting
consequence. The most remarkable feature of the relaxation
processes of the GMG performed with large $\delta p$ is that the
high fluctuations prevents quenching (see Fig. \ref{melting}).
This allows to provide a new framework to the annealing procedure
presented in Refs.\cite{bcp} and \cite{bcp-II} that resembles more
closely the traditional protocol of Ref.\cite{gelatt}.

The method presented in Ref.\cite{bcp} requires an iterative
procedure which involves a short evolution of the $N-$ agent
system and the subsequent elimination of all points accumulated
in the system. This is repeated until a moment in which the
distribution $P(p)$ remains stationary. With the present
interpretation of $\delta p$, a thermal annealing relaxation for
the GMG can be performed for the cases in which $\mu$ is
significantly different from 1/2. This new protocol can be assumed
to take place in episodes. In the first episode, relaxation is
allowed using a value of $\delta p$ that is large enough to
insure that equilibrium is reached and quenching is prevented.
The following episodes start from the equilibrium reached in the
preceeding one, and a new relaxation process is allowed with a
smaller value of $\delta p$ that is still large enough to avoid
the appearance of quenching. The process continues until a lower
bound of $\delta p$ is reached. Following this ``cooling"
protocol quenching never occurs, an absolute minimum of ${\cal
C}$ is obtained and the population remains strongly polarized.


\section{Conclusions}

In the present paper we provide and alternative description of the
dynamics of a system composed by many agents that play at the
GMG. This is given in terms of the optimization of a single
global magnitude, instead of doing it in terms independent
actions of the $N$ agents. We do this by studying the effect of
introducing a cost function ${\cal C}$ that is associated to the
second moment of the probability distribution of the size of the
attending parties.

We have proven that ${\cal C}$ has the relevant properties of an
internal energy. In equilibrium, it is a positive extensive quantity that
scales linearly with the number of agents $N$ and its minima correspond to
equilibrium configurations with a highly polarized population, as found
in the BAM or the GMG without quenching.

In addition, the deterministic dynamics that is derived from the descent
along the gradient of ${\cal C}$ leads the system to configurations that
have an equivalent polarization as that found with the traditional
stochastic updating of the BAM or the GMG.  This is a non trivial
equivalence between two completely different organization schemes of the
N-agent system.  On the one hand the gradient descent gives rise to a set
of coupled differential equations that represents a coordinated evolution
of all the agents as would be the result of the action of a ``central
planner" of the whole system.  On the other hand, within the GMG all the
agents act independently from each other adjusting their attendance
strategies with the purpose of optimizing their individual utilities.
Even though the two relaxation mechanisms are very different, the final
configurations of the system turn out to have equivalent features.

The definition of ${\cal C}$ in terms of the second moment of the
probability distribution of attending parties is reminiscent of
the many body Hamiltonian introduced in
Ref.\cite{ChalletMarsili} to cast a version of the MG into the
spin glass formalism. In the present case ${\cal C}$ can also be
considered as a many body Hamiltonian with one- and two-body
interactions in which the $N$ dynamic variables are the
attendance probabilities $p_i$'s, with $i=1,2 \dots N$.

The introduction of ${\cal C}$ and the associated relaxation process
allows to define a temperature parameter through a Langevin-like approach.
The value of $T$ remains associated to the ensemble average of the square
of the fluctuations of the attendance, scaled by the number of
agents.  Its introduction in ${\cal C}$ provides the proof that this
quantity, in thermal equilibrium, scales linearly with the size $N$ of the
system and therefore qualifies as an extensive parameter.

On the other hand, in order to be an intensive parameter, $T$ should be
independent of the size of the system.  This has been checked numerically
for the case of the GMG.  However finite size effects in the definition of
$T$ become negligible only for systems that are significantly larger than
the minimal ones that already display the self organization features and
that have spurred the popularity of the Minority Game.

Thermal fluctuations can be included in the dynamics that
corresponds to the descent along the gradient of ${\cal C}$. The
corresponding distributions $P(p)$ can readily be found and a
comparison can be made of $T$ with $\delta p$ involved in the
relaxation of the GMG or the BAM. A direct relationship can be
established between both parameters but only in the limit of
$\delta p \rightarrow 0$. We have also considered the dynamics of
the GMG with moderately large values of $\delta p$ when still the
divergence between the GMG and the thermal dynamics is not
important. A stochastic updating that involves large values of
$\delta p$ could be thought to be associated to irresolute or
badly informed agents that correct their attendance probabilities
performing {\em significant} changes in each correction.

The GMG relaxation for large values of $\delta p$ avoids
quenching even for $\mu$ significantly different from 1/2. This
fact, together with the thermal interpretation of $\delta p$
allows to cast the annealing procedure presented in Ref.\cite{bcp}
into the more traditional framework in which $T$ is progressively
reduced in successive epochs. This ``cooling" protocol could well
be assimilated to a succession of {\em learning episodes} of the
many agent system. In the first episodes in which agents have
little ``experience" and the information about the past is
scarce, all agents perform large amplitude - even random -
corrections. In the last episodes of the relaxation process, as
there is a richer information about the past history of the
system the agents perform finer corrections, the fluctuations are
smaller and the cost paid by a wrong attendance are also smaller.

The fact that on the one hand an extensive magnitude can be defined
playing the role of an internal energy, and that on the other, a
microscopic definition of the temperature can be made, opens the way to a
the full thermodynamic description of a system of N-agent performing a
GMG.  This amount to introduce a Gibbs distribution defined as $\Phi({\cal
C})= e^{-{\cal C}/T}/Z$, where $Z$ stands for the partition function.  All
thermodynamic functions should follow from this.

E.B. has been partially supported by CONICET of Argentina, PICT-PMT0051;H.C.
and R.P. were partially supported by EC grant ARG/b7-3011/94/27, Contract
931005 AR.

\end{multicols}


\begin{references}
\bibitem{Arthur2}  Amer. Econ. Assoc. Papers Proc. {\bf 84}, 406 (1994)

\bibitem{Challet}  D. Challet, Y.C. Zhang, Physica A{\bf 246}, 407 (1997);
Physica A{\bf 256}, 514 (1998)

\bibitem{Savit}  R. Savit, R. Manuca, R. Riolo, Phys. Rev. Lett. {82}, 2203
(1999)

\bibitem{Cavagna}  A. Cavagna, Phys. Rev. {\bf E59}, R3783 (1999)

\bibitem{Johnson}  N.F. Johnson, P.M. Hui, R. Jonson, T.S. Lo, Phys. Rev.
Lett. {82}, 3360 (1999)

\bibitem{bcp} E.Burgos, H. Ceva and R.Perazzo. Physica {\bf A294}, 539 (2001)

\bibitem{bcp-II} E.Burgos, H. Ceva and R.Perazzo. Phys Rev{\bf E}, in press

\bibitem{gelatt} S. Kirkpatrick, C.D. Gelatt and P.P. Vecchi, Science {\bf
220}, 671 (1983)


\bibitem{definicion} In Ref.\cite{bcp-II} the average of Eq.(\ref{costo})
is defined as ${\cal C}^2$ instead of ${\cal C}$ as we do here.  We prefer
the present notation because, as we prove later, this quantity, in
equilibrium, scales linearly with the size of the system as corresponds to
an extensive quantity.

\bibitem{Johnson2}  N.F. Johnson, D.J.T. Leonard, P.M. Hui, T.S. Lo,
cond-mat /9905939 v2. See also: N.F. Johnson, P.M. Hui, Dafang Zeng, C.W.
Tai, Physica {\bf A269}, 493 (1999)

\bibitem{bc}  E. Burgos, H.Ceva, Physica{\bf \ A284}, 489 (2000)

\bibitem{Goldberg}  A.Goldberg {\em Genetic Algorithms in Search,
Optimization, and Machine Learning}, Addison - Wesley (1989).

\bibitem{ChalletMarsili} D. Challet, M. Marsili Phys. Rev. {\bf E60},
R6271 (1999)

\bibitem{Ceva} H.Ceva, Physica{\bf \ A277}, 496 (2000)

\end{references}
\end{document}